\newlength{\absize}
\documentclass[12pt]{article}
\usepackage{latexsym}
\usepackage{amssymb}
\usepackage{epsf}
\usepackage{setspace}
\newcommand{\figsize}{\small}

\input prepictex
\input pictex
\input postpictex
\newdimen\tdim
\tdim=\unitlength
\def\stpltsmbl{\setplotsymbol ({\small .})}

\newbox\sru
\setbox\sru=\hbox{\beginpicture
\setcoordinatesystem units <\tdim,\tdim>
\stpltsmbl
\setquadratic
\plot
  0.0   0.0
  4.8   1.5
  7.5   5.0
  7.3   8.5
  5.0  10.0
  2.7   8.5
  2.5   5.0
  5.2   1.5
 10.0   0.0
/
\endpicture}
\def\springru #1 #2 *#3 /{\multiput {\copy\sru}  at
#1 #2 *#3 10 0 /}

\renewcommand{\bar}{\overline}

\newcommand{\un}{\mathcal{U}}
\newcommand{\bz}{\mathcal{BZ}}

\def\ltilde#1{\mathord{\mathop{\kern 0pt #1}\limits_{\sim\atop}}}

\begin{document}

\thispagestyle{empty}
\pagestyle{empty}
\renewcommand{\thefootnote}{\fnsymbol{footnote}}
\newcommand{\starttext}{\newpage\normalsize
 \pagestyle{plain}
 \setlength{\baselineskip}{3ex}\par
 \setcounter{footnote}{0}
 \renewcommand{\thefootnote}{\arabic{footnote}}
 }
\newcommand{\preprint}[1]{\begin{flushright}
 \setlength{\baselineskip}{3ex}#1\end{flushright}}
\renewcommand{\title}[1]{\begin{center}\LARGE
 #1\end{center}\par}
\renewcommand{\author}[1]{\vspace{2ex}{\Large\begin{center}
 \setlength{\baselineskip}{3ex}#1\par\end{center}}}
\renewcommand{\thanks}[1]{\footnote{#1}}
\renewcommand{\abstract}[1]{\vspace{2ex}\normalsize\begin{center}
 \centerline{\bf Abstract}\par\vspace{2ex}\parbox{\absize}{#1
 \setlength{\baselineskip}{2.5ex}\par}
 \end{center}}

\preprint{}
\title{Unparticle Physics}
\author{
 Howard~Georgi,\thanks{\noindent \tt georgi@physics.harvard.edu}
 \\ \medskip
Center for the Fundamental Laws of Nature\\
Jefferson Physical Laboratory \\
Harvard University \\
Cambridge, MA 02138
 }
\date{March 2007}
\abstract{I discuss some simple aspects of the low-energy physics of a
nontrivial scale invariant sector of an effective field theory --- physics
that cannot be described in terms of particles. I argue 
that it is important to take seriously the possibility that the
unparticle stuff described by such a theory might actually exist in our
world. I suggest a scenario in which some details of the production of
unparticle stuff can be calculated. I find that in the appropriate low
energy limit, unparticle
stuff with scale dimension $d_{\mathcal{U}}$ 
looks like a non-integral number 
$d_{\mathcal{U}}$ of invisible particles. Thus 
dramatic evidence for a nontrivial scale
invariant sector could show up experimentally in missing energy
distributions.}

\starttext

\setcounter{equation}{0}

Stuff with
nontrivial scale invariance in the infrared (IR)~\cite{Banks:1981nn} would be 
very unlike anything we have seen in our world. 
Our quantum mechanical world seems to be
well-described in terms of particles. 
We have a common-sense notion of what a particle is. 
Classical particles have
definite mass and therefore carry 
energy and momentum in a definite relation $E^2=p^2c^2+m^2c^4$. 
In quantum mechanics, this relation becomes the dispersion relation for the
corresponding quantum waves with the mass fixing the low-frequency
cut-off, $\omega^2=c^2k^2+m^2c^4/\hbar^2$. 

Scale invariant stuff cannot have a definite mass unless that mass is zero. 
A scale transformation
multiplies all dimensional quantities
by a rescaling factor raised to the mass dimension so
a nonzero mass is not scale invariant. 
A free massless particle is a simple
example of scale invariant stuff because the zero mass is unaffected by
rescaling. But quantum field theorists have long 
realized that there are more interesting possibilities --- theories in which
there are fields that get multiplied by fractional powers of the rescaling
parameter.\footnote{See for example \cite{Wilson:1970pq}.} 
The standard model does not 
have the property of scale invariance. Many of our particles have definite
nonzero masses.\footnote{\label{broken}It
could be that the high energy is scale invariant, but
that the scale invariance is broken at or above the electroweak scale. That
is {\bf not} what I am discussing. It leads to ordinary particles.} But
there could be a sector of the 
theory, as yet unseen, that is exactly
scale invariant and very weakly interacting
with the rest of the standard model.\footnote{I will make this precise below.}
In such an interacting scale invariant
sector in four space-time dimensions, there are
no particles because there can be no particle states with 
a definite nonzero mass. 
Scale invariant stuff, if it exists, is made of unparticles. 

But what does this mean? It is clear what scale invariance is in the
quantum field theory. Fields can scale with fractional dimensions. 
Indeed, much beautiful theory is devoted 
working out the
structure of these theories.\footnote{The huge literature intersects with
supersymmetry (for a review see \cite{Intriligator:1995au}), with
string theory (for a review see \cite{Aharony:1999ti}) and particularly
with the
AdS/CFT conjecture (for a review see \cite{Maldacena:2003nj}). } But 
what would scale invariant unparticle stuff actually look like
in the laboratory? In spite of all we know about the correlation functions of
conformal fields in Euclidean space,
it is a little
hard to even talk about the physics of something so different from our
familiar particle theories.
It does not seem {\it a priori\/} very likely that such different stuff
should exist and have remained hidden.
But this is no reason to assume that it is impossible.
We should determine {\bf experimentally} whether 
such unparticle stuff actually exists. But
how will we know if it we see it? That is one of the 
questions I address in this note.

I discuss a simple scenario in which we can say something
simple and unambiguous about what unparticles look like. 
The tool I use to say something quantitative
about unparticle physics is effective field theory~(see for example
\cite{Georgi:1994qn}).  
The idea is that while 
the detailed physics of a theory
with a nontrivial scale invariant infrared fixed point is thoroughly
nonlinear and complicated, the low
energy effective field theory, while very strange, is very simple because of
the scale invariance. We can use this to understand 
what the interactions of unparticles
with ordinary matter look like  in an appropriate limit. 
Parts of what I have to say
are well understood by many experts in scale invariant field
theories.\footnote{See, for example, \cite{Aharony:1999ti}.
But note that one
reason that it is difficult to extract unparticle physics from the
beautiful formal works on conformal theory is that these papers often have
in mind the scheme described in footnote~\ref{broken}.} 
I hope to make it 
common knowledge among phenomenologists and experimenters.
My goal here is not to do serious phenomenology myself, but 
rather to describe very clearly a physical situation in which
phenomenology is possible in spite of the essential strangeness of
unparticle theories. And while my motivation is primarily just theoretical
curiosity, the scheme I discuss could very well be a component of the
physics above the TeV scale that will show up at the LHC. To my
mind, this would be a much more striking discovery than the
more talked about possibilities of SUSY or
extra dimensions. SUSY is more new particles. 
From our 4-dimensional point of view until we see black holes or
otherwise manipulate gravity, finite extra
dimensions are just a metaphor.\footnote{Infinite extra dimensions,
however, can have unparticle-like behavior. See \cite{Randall:1999vf}.} 
Again what we see is 
just more new particles. We would be
overjoyed and fascinated to see these new particles and eventually patterns
might emerge that show the beautiful theoretical
structures they portend. But I will argue that
unparticle stuff with nontrivial scaling would
astonish us immediately.

Here is the scheme. The very high energy
theory contains the fields
of the standard model and the fields of a theory with a
nontrivial IR fixed point, which we will call $\bz$ (for Banks-Zaks) fields.
The two sets interact through the exchange of particles with a 
large mass scale $M_{\un}$. Below the scale $M_{\un}$, there are
nonrenormalizable couplings involving both standard model
fields and Banks-Zaks
fields suppressed by powers of $M_{\un}$. These have the generic form
\begin{equation}
\frac{1}{M_{\un}^k}O_{sm}O_{\bz}
\label{sandbz}
\end{equation} 
where $O_{sm}$ is an operator with mass dimension $d_{sm}$ built out of
standard model fields and  
$O_{\bz}$ is an operator with mass dimension $d_{\bz}$ built out of
$\bz$ fields.
The renormalizable couplings of the $\bz$ fields then cause
dimensional transmutation as scale-invariance in the $\bz$ sector 
emerges at an energy
scale $\Lambda_{\un}$.
In the effective theory below the scale $\Lambda_{\un}$ the $\bz$ operators
match onto unparticle operators, and the interactions
of (\ref{sandbz}) match onto interactions of the form
\begin{equation}
\frac{C_{\un}\,\Lambda_{\un}^{d_{\bz}-d_{\un}}}{M_{\un}^k}O_{sm}O_{\un}
\label{sandu}
\end{equation} 
where $d_{\un}$ is the scaling dimension of the unparticle operator
$O_{\un}$.\footnote{For now
we assume for simplicity of presentation that $O_{\un}$
is a Lorentz scalar. See (\ref{tdofs}).} 
The constant $C_{\un}$ is a coefficient function. 
We are interested in the operators of the lowest
possible dimension, which have the largest effect in the low energy
theory, so we will assume that $O_{\un}$ is one such.
The effective field theory interaction (\ref{sandu}) is a good starting
point in our search for unparticle stuff, for two reasons. Because the $\bz$
fields decouple from ordinary matter at low energies, the interaction
(\ref{sandbz}) should not effect the IR scale invariance of the unparticle.
And (\ref{sandbz}) seems likely to be
allowed experimentally for sufficiently large
$M_{\un}$.
If $M_{\un}$ is large enough, the unparticle stuff
just doesn't couple strongly enough to ordinary stuff to have been
seen. What happens as we lower $M_{\un}$ or
raise our machine energy and this peculiar
stuff can be produced by interactions of ordinary particles?

If the IR fixed
point is perturbative, we may be able 
to calculate the $d_{\un}$s and $C_{\un}$s. 
But typically the
matching from the $\bz$ physics to the unparticle physics will be a
complicated strong interaction problem, like the matching from the physics
of high-energy QCD onto the physics of the low-energy hadron states. In
that case, we should be able to estimate these constants very roughly by
including the appropriate geometrical factors (powers of
$4\pi$ and that sort
of thing - we will return to this below), 
but detailed calculation will be impossible. 

Now we can ask what physics this produces in the low energy theory below
$\Lambda_{\un}$. We expect that the virtual effects of
fields with nontrivial scaling will produce odd forces. But here I
consider what it looks like to actually produce the unparticle stuff.
The most important effects will be those that involve only one
factor (in the amplitude) of the small parameter in (\ref{sandu}),
\begin{equation}
\frac{C_{\un}\,\Lambda_{\un}^{d_{\bz}-d_{\un}}}{M_{\un}^k}
\end{equation}
from a single insertion of the interaction (\ref{sandu}) in some standard
model process. The result will be the production of unparticle stuff, which
will contribute to missing energy and momentum. To calculate the
probability distribution for such a process, we need to know the density of
final states for unparticle stuff.
In the low energy theory described above, this is
constrained by the
scale invariance. Consider the vacuum matrix element
\begin{equation}
\left\langle0\right|O_{\un}(x)\,O_{\un}^\dagger(0)\left|0\right\rangle
=\int\,e^{-ipx}\,
\left|\left\langle0\right|O_{\un}(0)\left|P\,\right\rangle\right|^2\,
\rho\left(P^2\right)\,\frac{d^4P}{(2\pi)^4}
\label{vacuumme}
\end{equation}
where $\left|P\,\right\rangle$ is the unparticle state with 4-momentum
$P^\mu$ produced from the vacuum by $O_{\un}$.
Because of scale invariance, the matrix element (\ref{vacuumme}) scales
with dimension $2d_{\un}$, which requires that
\begin{equation}
\left|\left\langle0\right|O_{\un}(0)\left|P\,\right\rangle\right|^2\,
\rho\left(P^2\right)
=A_{d_{\un}}\,\theta\left(P^0\right)
\,\theta\left(P^2\right)\,\left(P^2\right)^{d_{\un}-2}
\label{dofs}
\end{equation}
This is the appropriate phase space for unparticle stuff.
(\ref{dofs}) should remind you of the phase space for $n$ massless
particles,\footnote{The left hand side has an extra $(2\pi)^4$ compared
to the definition in the particle date book.}
\begin{equation}
(2\pi)^4\delta^4\left(P-\sum_{j=1}^{n}p_j\right)
\prod_{j=1}^{n}
\,\delta\left(p_j^2\right)\,\theta\left(p_j^0\right)\,\frac{d^4p_j}{(2\pi)^3}
=A_n\,\theta\left(P^0\right)
\,\theta\left(P^2\right)\,\left(P^2\right)^{n-2}
\label{nparticle}
\end{equation}
where
\begin{equation}
A_n=\frac{16\pi^{5/2}}{(2\pi)^{2n}}
\,\frac{\Gamma(n+1/2)}{\Gamma(n-1)\,\Gamma(2n)}
\label{an}
\end{equation}
The zero
in $A_n$ for $n=1$ together with the pole in $P^2$ reproduce the
$\delta(P^2)$ in 1-particle phase space if the limit $n\to1$ is approached
from above
\begin{equation}
\mathop{\lim}\limits_{\epsilon\to0+}
\frac{\epsilon\,\theta(x)}{x^{1-\epsilon}}=\delta(x)
\end{equation}

Thus we can describe the situation concisely as follows:
\begin{equation}
\begin{minipage}{.65\textwidth}{{\bf Unparticle
stuff with scale dimension $d_{\un}$ looks like a non-integral number 
$d_{\un}$ of invisible particles.}}
\end{minipage}
\label{mantra}
\end{equation}
In fact, 
we may as well identify the $A$ in (\ref{dofs}) with the $A$ in
(\ref{an}), and thus adopt $(\ref{an})$ for non-integral $n$ as the
normalization for $A_{d_{\un}}$. This is purely conventional
because a different definition could be absorbed in the coefficient
function $C_{\un}$ in (\ref{sandu}),
but this choice fixes the normalization of the field $O_{\un}$ in a
way that incorporates the geometrical factors that go with dimensional
analysis, although the combinatoric factors may be wildly wrong. 

To illustrate the procedure in a realistic situation 
consider the decay $t\to u+ \un$ of a $t$ quark into a $u$ quark plus
unparticles of scale dimension $d_{\un}$
from the coupling\footnote{Chosen for simplicity rather than interest!}
\begin{equation}
i\,\frac{\lambda}{\Lambda^{d_{\un}}}
\,\bar u\,\gamma_\mu(1-\gamma_5)\,t\,\partial^\mu O_\un
+\mbox{h.c.}
\label{tuunl}
\end{equation}
where the constant 
$\lambda$ 
\begin{equation}
\lambda=\frac{C_{\un}\,\Lambda_{\un}^{d_{\bz}}}{M_{\un}^k}
\end{equation}
(which in this particular case is dimensionless)
contains most of the factors from the matching
onto the low energy theory. 
We can ignore the mass of the $u$ quark, so the
final state densities are
\begin{equation}
d\Phi_{u}(p_{u})=2\pi\,\theta\left(p_{u}^0\right)
\,\delta\left(p_{u}^2\right)
\end{equation}
\begin{equation}
d\Phi_\un(p_\un)=A_{d_{\un}}\,\theta\left(p_\un^0\right)
\,\theta\left(p_\un^2\right)\,\left(p_\un^2\right)^{d_\un-2}
\end{equation}
The way the phase space factors compose in my normalization is
\begin{equation}
d\Phi(P)=\int\,(2\pi)^4\delta^4\left(P-\sum_jp_j\right)
\prod_jd\Phi(p_j)\frac{d^4p_j}{(2\pi)^4}
\end{equation}
and the differential decay rate is
\begin{equation}
d\Gamma=\frac{|\mathcal{M}|^2}{2M}\,d\Phi(P)
\end{equation}
where $\mathcal{M}$ is the invariant matrix element. Suitably averaged over
initial spin and summed over final spin this gives
\begin{equation}
\frac{d\Gamma}{dE_{u}}=\frac{A_{d_{\un}}
m_t^2\,E_{u}^2\,|\lambda|^2}{2\pi^2\,\Lambda_\un^{2d_{\un}}}
\,\frac{\theta\left(m_t-2E_{u}\right)}
{\left(m_t^2-2m_tE_{u}\right)^{2-d_\un}}
\end{equation}
We are primarily interested in the shape as a function of $E_u$, so we will
plot $d\ln\Gamma/dE_u$ which has the simple form
\begin{equation}
\frac{1}{\Gamma}\frac{d\Gamma}{dE_{u}}=
4\,d_{\un}(d_{\un}^2-1)\,(1-2E_u/m_t)^{d_{\un}-2}E_u^2/m_t^2
\end{equation}
The result is shown in figure~\ref{fig-1}.
{\figsize\begin{figure}[htb]
$$\beginpicture
\setcoordinatesystem units <\tdim,\tdim>
\put {{\epsfxsize=290\tdim \epsfbox{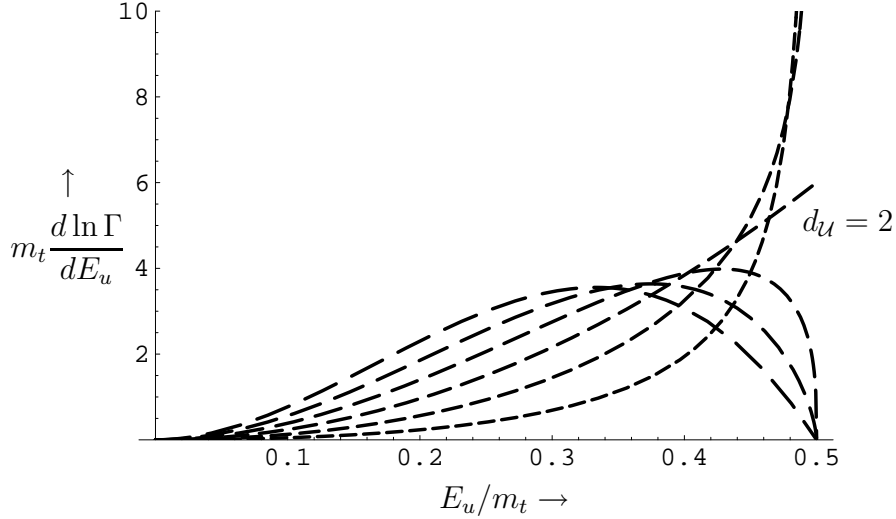}}} at 0 0
\put {$E_u/m_t\to$} at 10 -100
\put {$d_\un=2$} at 140 5
\put {$\displaystyle{m_t}\frac{d\ln\Gamma}{dE_u}$}
at -155 -5
\put {$\uparrow$} at -155 20
\endpicture$$
\caption{\small\sf\label{fig-1}$d\ln\Gamma/dE_u$ versus $E_u$ in units of
$m_t$ with $d_{\un}=j/3$ for $j=4$ to $9$. The dashes get longer as $j$
increases. }\end{figure}}
As $d_{\un}\to1$ from above, $d\ln(\Gamma)/dE_u$ becomes more peaked at
$E_u=m_t/2$, matching smoothly unto the kinematics of a 2-particle decay in
the limit, as expected from the general principle (\ref{mantra}).
Obviously, for higher $d_{\un}$ 
the shape depends sensitively on $d_u$, but at least for
$d_{\un}$ in this range, the calculation appears to make sense. The kind of
peculiar distributions of missing energy that we see in figure~\ref{fig-1}
may allow us to discover unparticles experimentally! The particular
operator (\ref{tuunl}) is flavor changing, and thus may be suppressed by
small and unknown flavor factors. But a similar analysis applies to
scattering processes due to flavor conserving operators.
The most interesting straightforward things to look at I believe are the
collider phenomenology of 
\begin{equation}
q+\bar q\to G+\un\quad\mbox{and}\quad q+G\to q+\un
\end{equation}
from the operators 
\begin{equation}
\frac{C_{\un}\,\Lambda_{\un}^{k+1-d_{\un}}}{M_{\un}^k}\;
\bar q\,\gamma_\mu\,q\,O^\mu_{\un}
\label{qandu}
\end{equation} 
where $q$ is a left- or right-handed quark, 
and the LEP constraints on the
operators
\begin{equation}
\frac{C_{\un}\,\Lambda_{\un}^{k+1-d_{\un}}}{M_{\un}^k}\;
\bar e\,\gamma_\mu(1\pm\gamma_5)\,e\,O^\mu_{\un}
\label{eandu}
\end{equation}
where the unparticle operator is hermitian and transverse,
\begin{equation}
\partial_\mu O^\mu_{\un}=0
\label{transverse}
\end{equation}
The calculation of matrix elements goes the same way except for the tensor
structure. For example
\begin{equation}
\left\langle0\right|O^\mu_{\un}(0)\left|P\,\right\rangle\,
\left\langle P\right|O^\nu_{\un}(0)\left|0\,\right\rangle\,
\rho\left(P^2\right)
=A_{d_{\un}}\,\theta\left(P^0\right)
\,\theta\left(P^2\right)\,
\left(-g^{\mu\nu}+P^\mu P^\nu/P^2\right)
\,\left(P^2\right)^{d_{\un}-2}
\label{tdofs}
\end{equation}
Also amusing is
\begin{equation}
G+G\to G+\un
\end{equation}
from the gluon operators
\begin{equation}
\frac{C_{\un}\,\Lambda_{\un}^{k-d_{\un}}}{M_{\un}^k}\;
G_{\mu\nu}G^{\mu\nu}\,O_{\un}
\quad\mbox{and}\quad
\frac{C_{\un}\,\Lambda_{\un}^{k-d_{\un}}}{M_{\un}^k}\;
G_{\mu\lambda}{G_\nu}^{\lambda}\,O^{\mu\nu}_{\un}
\label{gandu}
\end{equation} 

I have argued in this brief note that unparticle stuff with nontrivial
scaling dimension might exist in our world, and that
up to constants associated with the binding of massless matter into
unparticles, we can predict interesting features of unparticle
production that serve as experimental tests of this crazy possibility.
Let me close with some remarks.
\begin{itemize}
\item Many remarkable things are 
known about the scale and conformal invariant theories in 2
dimensions (see for example \cite{Ginsparg:1988ui}). It is not
clear to me what 2D results translate into 4D because the phase space in 2D
is so constrained. But there are certainly 
consequences for condensed matter physics,
where conformal structures do exist (see for example \cite{Cardy:2003zr}).
\item The connection between operator scaling dimension in a cft and missing
energy distributions was made for 
ordinary particles with integral scaling dimension
in \cite{Arkani-Hamed:2000ds}. 
\item The effective field theory picture above 
assumes that the unparticle fields do
not carry the standard model gauge interactions. It would be interesting to
try to relax this, but I have no idea whether it is possible.
\item In (\ref{sandu}), (\ref{qandu}), (\ref{eandu}) and (\ref{gandu}),
we assumed that the unparticle operator is a bosonic field.
Fermionic fields are possible if the standard model
fields include fermions and bosons with the same gauge couplings, as in
SUSY, or if one can makes sense of unparticle fields with standard model
gauge quantum numbers.
\item If unparticles exist, their cosmological consequences
should be investigated. It should be possible to use effective field theory
to understand how low energy unparticles behave in the universe today. But
additional tools may be required to understand how they got there from the
hot big bang.
\item I had hoped briefly to
make sense of unparticles with $d_{\un}<1$.
However, in the calculation leading to
figure~\ref{fig-1} 
the differential decay rate into unparticles with $d_{\un}<1$
has a non-integrable singularity as $E_{\un}\to0$, suggesting that the vacuum
might be unstable. This is in accord with the general theorem in
\cite{Mack:1975je} that such fields are not possible in a unitary theory
(one of many important contributions by this author to the subject of
conformal field theory).
\end{itemize}

{\it Note Added:\/} While this paper was being reviewed, a number of papers
have appeared applying its ideas.~\cite{Georgi:2007si,
Cheung:2007ue,
Luo:2007bq,
Chen:2007vv,
Ding:2007bm,
Liao:2007bx,
Aliev:2007qw,
Catterall:2007yx,
Li:2007by}
Several of these
discuss an extension of the ideas discussed in
this paper to the interference between virtual unparticles in the
$s$-channel and standard model amplitudes.

{\it Acknowledgments:\/}
I am grateful to Nima Arkani-Hamed, Tom Banks, 
Spencer Chang, Ann Nelson, Lisa
Randall and Edward Witten for comments on the manuscript.
This research is supported in part by
the National Science Foundation under grant PHY-0244821.


\providecommand{\href}[2]{#2}\begingroup\raggedright\endgroup

\end{document}